\documentclass[ aip, amsmath,amssymb, reprint, ]{revtex4-2}

\usepackage{hyperref}
\usepackage{graphicx}
\usepackage{bm}
\usepackage[utf8]{inputenc}
\usepackage[T1]{fontenc}
\usepackage{newtxtext}
\usepackage{newtxmath}

\def\eg{\emph{e.g.}}

\def\Hm{\hat H^{\scriptscriptstyle <}}

\def\Am{A^{\scriptscriptstyle <}_\lambda}
\def\Ap{A^{\scriptscriptstyle >}_\lambda}

\def\JJ{\bm{J}}
\def\FF{\bm{F}}
\def\rr{\bm{r}}
\def\kk{\bm{k}}
\def\ff{\bm{f}}
\def\jj{\bm{\jmath}}

\begin{document}

\title{Green and Kubo forge the arrow of time}
\author{Stefano Baroni}
 \affiliation{SISSA -- Scuola Internazionale Superiore di Studi Avanzati, 34136 Trieste, Italy}
 \affiliation{CNR -- Istituto dell'Officina dei Materiali, SISSA unit, 34136 Trieste, Italy}
 \email{baroni@sissa.it}

\date{\today}

\begin{abstract}
Transport theory describes the response of a macroscopic current to a thermodynamic force, thus producing entropy and apparently violating time-reversal symmetry. In this note I report a pedagogical derivation of the Green-Kubo formula for transport coefficients, based on elementary equilibrium statistical mechanics and static response theory, that highlights the intrinsically dynamical nature of this formula and showcases the relation between the apparent breach of time-reversal symmetry and the non-commutativity of the low-frequency / low-wavevector limits of the conserved-density susceptibilities, from which the formula can be established.
\end{abstract}

\maketitle

\section{Introduction}
Heat flows from warmth to coolth as time flows from the past to the future. As a matter of fact, the flow of energy (or of any conserved quantity such as mass or charge, for that matter) is the most basic mechanism of entropy production and a fundamental manifestation of the arrow of time.

At thermodynamic equilibrium, the intensive variables ($\alpha$) conjugate to the extensive and conserved arguments ($A$) of the micro-canonical entropy, $\alpha=\frac{\partial S(A)}{\partial A}$, are constant across the volume ($\Omega$) of a macroscopic system. When thermodynamic equilibrium holds only locally, intensive variables may weakly depend on time and position, thus determining a flow of the extensive conserved quantities conjugate to them, so as to restore global equilibrium. Transport theory describes how the current densities of the conserved quantities, $\bm\jmath(\bm r)$ (\emph{conserved currents}, in short), respond to the gradients of their conjugate intensive variables (the \emph{thermodynamic forces}), $\bm f(\bm r)=\nabla\alpha(\bm r)$. For the sake of definiteness, if the entropy is thought to be a function of energy, volume, and number of molecules of each molecular species, $A=\{E,\Omega,N_i\}$, the corresponding intensive variables are $\alpha=\left\{\frac{1}{T}, \frac{p}{T}, -\frac{\mu_i}{T}\right \}$, where $T$, $p$, and $\mu_i$ are temperature, pressure, and the various chemical potentials, the latter possibly including the effects of external potentials. For future reference, let us define a \emph{(conserved) flux} as the macroscopic average of a (conserved) current, $\bm J=\frac{1}{\Omega}\int \bm \jmath(\bm r)d\bm r$, and analogously the macroscopic force as  $\bm F=\frac{1}{\Omega}\int \bm f(\bm r)d\bm r$. It is also worth noting that macroscopic averages are the long-wavelength (small wavevector) limits of the corresponding Fourier transforms,
defined as 
\begin{equation}
    \tilde g(\kk)=\frac{1}{\Omega}\int g(\rr)e^{-i\kk\cdot\rr}d\rr.
\end{equation}
One has therefore: $\bm J=\tilde{\bm \jmath}(0)$ and $\bm F=\tilde{\bm f}(0)$ 

At equilibrium both forces and currents vanish. When thermodynamic forces are small, the conserved fluxes are proportional to them:
\begin{equation}
    \bm J = \sigma \bm F. \label{eq:Onsager}
\end{equation}
Eq. \eqref{eq:Onsager} states that the response of a quantity that is odd with respect to time reversal (the current) is proportional to another that is instead even (the force), thus violating time-reversal symmetry and determining an increase of entropy.

Let a system of volume $\Omega$ be partitioned into $n$ subvolumes, $\{ \Omega_{1}, \Omega_{2},\cdots\Omega_{n}\}$. The rate of change of the total entropy is:
\begin{align} \label{eq:EntropyProduction}
    \begin{aligned}
        \frac{dS(\Omega,t)}{dt} & =\sum_{i}\frac{\partial S(\Omega_{i},t)}{\partial A_{i}}\frac{dA_{i}}{dt}\\
        & =\sum_{i}\alpha_{i}\dot{A}_{i}.
    \end{aligned}
\end{align}
In the continuous limit, the sum over the subvolumes can be replaced by an integral and Eq. \eqref{eq:EntropyProduction} be cast into the form:
\begin{equation} \label{eq:EntropyProductionContinuum}
        \frac{dS(\Omega,t)}{dt} = \int_{\Omega}\alpha(\bm{r},t) \dot{a}(\bm{r},t) d\bm{r},
\end{equation}
where $a(\bm r)$ is the density of the conserved quantity $A$ (a \emph{conserved density}), satisfying the continuity equation:
\begin{equation}
 \dot{a}(\bm{r},t)+\nabla\cdot\bm{\jmath}(\bm{r},t)=0,\label{eq:continuity}
\end{equation}
which is the defining relation of any locally conserved extensive quantity. Using Eq. \eqref{eq:continuity} to express the time derivative of the density in terms of the divergence of the current and integrating by parts, Eq. \eqref{eq:EntropyProductionContinuum} can be written as:
\begin{align}
    \begin{aligned}
        \frac{dS(\Omega,t)}{dt} & =-\int_{\Omega}\alpha(\bm{r},t)\nabla\cdot\bm{\jmath}(\bm{r},t)d\bm{r}\\
    & =\int_{\Omega}\nabla\alpha(\bm{r},t)\cdot\bm{\jmath}(\bm{r},t)d\bm{r}.
    \end{aligned}
   \label{eq:SSource}
\end{align}
The integrand in Eq. \eqref{eq:SSource} can be interpreted as an entropy source, $ \dot s(\bm r,t)=\nabla\alpha(\bm{r},t)\cdot\bm{\jmath}(\bm{r},t)$. By combining Eq. \eqref{eq:SSource} with Eq. \eqref{eq:Onsager}, which in the long-wavelength limit can be assumed to hold locally, and requiring that the rate of change of the entropy is positive, one obtains the condition that that transport coefficients are positive: $\sigma>0$. Heat flows indeed from warmth to coolth!

\begin{figure}[t!]
    \centering
    \includegraphics[width=0.95\columnwidth]{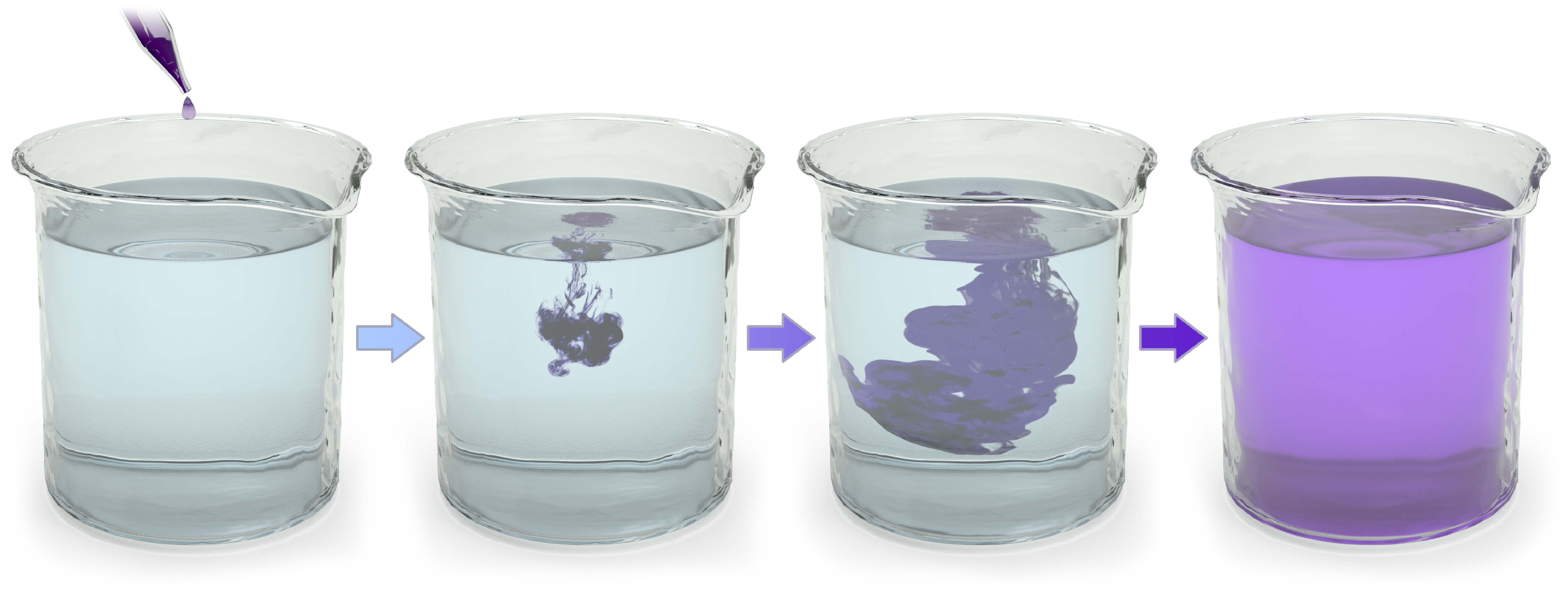}
    \caption{A drop of dye is dripped into a solvent, dispersing through the vessel until it is evenly spread across it. Credit: Bruce Blaus \cite{BruceBlaus}} \label{fig:Diffusion}
\end{figure}

For the sake of definiteness, I will explicitly address the case of diffusion, illustrated in Fig. \ref{fig:Diffusion}. Here, the conserved quantity is represented by the number of dye molecules, $N$, whose density and current density are  denoted by $n$ and $\bm\jmath$, respectively. If diffusion takes place at constant temperature, the thermodynamic force acting on the system is $\bm f = -\frac{1}{T} \nabla \mu$. In order to simplify the notation, in the following I will redefine the \emph{force} as the gradient of the (chemical) potential, $\bm f \rightarrow T\bm f = -\nabla\mu$, and incorporate the inverse temperature that appears in its proper definition into the transport coefficient: $\sigma\rightarrow \sigma/T$. In the absence of external forces, the gradient of the chemical potential is proportional to the gradient of the density and the force reads therefore: $\bm f = - \frac{\partial\mu}{\partial n} \nabla n$. In the long-wavelength limit, Eq. \eqref{eq:Onsager} holds locally and can thus be turned into Fick's law:\cite{Fick1855} 
\begin{equation}
  \bm{\jmath}=-D\nabla n,\label{eq:fick}
\end{equation}
where $D=\sigma \frac{\partial\mu}{\partial n}$ is the dye's diffusivity. By combining Fick's law with the continuity equation for the dye number density, Eq. \eqref{eq:continuity}, we finally arrive at the diffusion equation:
\begin{equation} \label{eq:diffusion}
    \frac{\partial n(\bm r,t)}{\partial t} = D\Delta n(\bm r,t).
\end{equation}
The solution of this equation is particularly simple in reciprocal space, where it is turned into a linear, first-order, ordinary differential equation for the Fourier transform of the number density, $\tilde n(\kk)$: $\dot{\tilde n}=-Dk^2\tilde n$. The time evolution of $\tilde n$ is then:
\begin{equation} \label{eq:n(k,t)}
   \tilde n(\bm k,t)= \tilde n(\bm k,0) e^{-D k^2 t}.
\end{equation}

\section{Linear response} \label{sec:LinearResponse}
When the force acting on the system is mechanical, an explicit expression for the $\sigma$ transport coefficient in Eq. \eqref{eq:Onsager} can be derived using (classical) Hamiltonian perturbation theory. Similar arguments could be used in the quantum-mechanical case, but here I restrict myself to the classical regime. The diffusion process discussed in the introduction can be modeled mechanically by supposing that the concentration inhomogeneity of the dye is determined by a static external force, whose effect is to confine it. This force is assumed to have been active from the distant past and then switched off at a time $t=0$. More generally, when the system experiences a non-mechanical force (such as a temperature or chemical-potential gradient), the definition of a suitable mechanical proxy would allow us to treat it by Hamiltonian perturbation theory, as explained \eg\ in Ref. \onlinecite{Luttinger1964}.

The states of a classical system of $N$ particles are identified by the coordinates of a point in phase space, $\Gamma=\{q,p\}$, where $q\doteq \{\bm{R}_{n}\}$ and $p\doteq \{\bm{P}_{n}\}$ are the sets of coordinates and momenta of the particles, which satisfy Hamilton's  equations of motion, $ {\dot{q}}_{t}  = \frac{\partial H^\circ}{\partial p}$ and $ {\dot{p}}_{t} = -\frac{\partial H^\circ}{\partial q}$, where
\begin{align}
    H^{\circ}(\Gamma)=\sum_{n}\frac{(\bm{P}_{n})^{2}}{2M_{n}} + U(\bm{R}_{1},\bm{R}_{2},\cdots\bm{R}_{N}),\label{eq:unperturbed_H}
\end{align}
is the system's (unperturbed) Hamiltonian, $\{M_{n}\}$ being the particles' masses and $U$ a generic translationally invariant many-body potential. Classical ``observables'' are functions defined in the system's phase space. I adopt the convention that a caret, as in $\hat A$, indicates an implicit dependence on the system's phase-space coordinates, $\Gamma=\{q,p\}$: $\hat A = A(\Gamma)$. When neither a hat nor the phase-space argument are present, $A$ indicates the expectation of $\hat A$ over some suitably defined phase-space probability density, $\mathsf P(\Gamma)$: $A = \langle \hat{A} \rangle \doteq \int A(\Gamma) \mathsf P(\Gamma) d\Gamma$. Sometimes, we will need to indicate explicitly the implicit time dependence of a phase-space variable (no pun intended). In this case, I will mean the phase-space function $\hat A(t)=A(\Gamma,t)\doteq A(\Gamma_t)$, as a function of time and of the initial condition, $\Gamma=\Gamma_0$, of a phase-space trajectory, $\Gamma_t=\{q_t,p_t\}$, as determined by Hamilton's equations of motion.

When a system is subject to a (static) perturbation,
\begin{align} \label{eq:static-perturbation}
    \hat{H}_\lambda=\hat{H}^{\circ}+\lambda\hat{V},
\end{align}
for small enough $\lambda$ the equilibrium expectation value of an observable depends linearly on the strength of the perturbation:
\begin{equation} \label{eq:chiAV0}
    A' = \langle\hat{A}\rangle_\lambda - \langle\hat{A}\rangle_0 \approx -\lambda \beta C_{AV}. 
\end{equation}
In this equation, $\langle\cdot \rangle_\lambda$ indicates that the expectated value is computed with respect to the equilibrium distribution corresponding to the given value of $\lambda$, $\mathsf P_\lambda$, which in the canonical ensemble reads:
\begin{equation} \label{eq:A_lambda}
    \begin{aligned}
        \langle \hat A\rangle_\lambda &= \int A(\Gamma) \mathsf P_\lambda(\Gamma) d\Gamma \\
        &= \frac{\int e^{-\beta H_\lambda(\Gamma)}A(\Gamma)d\Gamma}{\int e^{-\beta H_\lambda(\Gamma)}d\Gamma},
    \end{aligned}
\end{equation}
and the correlation function $C_{AV}$ is defined as:
\begin{equation} \label{eq:chiAV}
    C_{AV} = \langle \hat{A}\hat{V}\rangle_0-\langle \hat{A}\rangle_0\langle \hat{V}\rangle_0. \end{equation}
Eq. \eqref{eq:chiAV0} can be easily derived by expanding the expected value of the observable $\hat A_\lambda$, Eq. \eqref{eq:A_lambda}, in powers of $\lambda$ and retaining only terms linear in $\lambda$.

We now suppose that a static perturbation has been switched on in the distant past and that the system is then let equilibrate until at a certain time, $t=0$, it is switched off again. The time-dependent Hamiltonian would then read:
\begin{equation} \label{eq:Hm(t)}
    \Hm (t)=\hat H^\circ +\lambda\theta(-t) \hat V,
\end{equation}
where $\theta$ is the Heaviside step function. For $t>0$ the system evolves with the unperturbed Hamiltonian. The time evolution of the expected value of the $\hat A$ observable, $A(t)$, is therefore the expected value of $A(\Gamma_t)$ with respect to the initial conditions of the $\Gamma_t$ trajectory, which are distributed according to the perturbed equilibrium phase-space density. Formally, $\hat{A}(t)$ can be thought of as a static variable that depends on the initial conditions, $\Gamma_0$, of the trajectory, $\Gamma_t$, and parametrically on time: $\hat A(t)=A(\Gamma_t)=A(\Gamma_0,t)$. Static linear-response theory then applies and the expectation value of $\hat{A}(t)$ as a function of the strength $\lambda$ of the perturbation is:
\begin{equation} \label{eq:relax1}
    \Am(t) \doteq A_\lambda(t)-A_0 =
    \begin{cases}
      ~- \lambda\beta C_{AV}(0)& \text{if } t \le 0 \\
      ~- \lambda\beta C_{AV}(t)& \text{if } t>0 \\
    \end{cases},
\end{equation}
where, assuming that the expected values of the $\hat A$ or $\hat V$ observables vanish at equilibrium, the time correlation function $C_{AB}(t)$ is defined as in Eq. \eqref{eq:chiAV}:
\begin{equation}
  \begin{aligned}
    C_{AV}(t) & =\left\langle \hat{A}(t)\hat{V}\right\rangle \\
    & \doteq\int A(\Gamma_{t}) V(\Gamma_{0}) \mathsf{P}_0(\Gamma_{0}) d\Gamma_{0}.
  \end{aligned}
\end{equation}
Note that, if $\hat{A}$ and $\hat{V}$ become uncorrelated at large time lags, Eq. \eqref{eq:relax1} indicates that at large times the value of $\hat{A}$ relaxes to equilibrium $(\Am(+\infty)=A_0)$, as one expects.

Eq. \eqref{eq:relax1} expresses the regression of the departure from equilibrium of an observable caused by a perturbation of strength $\lambda$. If we express this strength in terms of the amplitude of the initial ($t=0$) distortion, $\lambda= -\frac{A'}{\beta \langle\hat A\hat V\rangle}$, see Eqs. (\ref{eq:chiAV0}-\ref{eq:chiAV}), one gets:
\begin{align}
  \Am(t) = \Am(0) \frac{\langle \hat A(t) \hat V\rangle}{\langle \hat A \hat V\rangle}. \label{eq:regression}
\end{align}
This equation expresses the fact that, in the linear regime, the time dependence of the regression to equilibrium of a disturbance induced by an external perturbation is the same as that of a spontaneous fluctuation. This fact is known as \emph{Onsager's regression hypothesis}.\cite{onsagerI1931,*onsagerII1931}

If the perturbation in Eq. \eqref{eq:Hm(t)} were switched on, rather than off, at time $t=0$, by linearity the response would read:
\begin{equation} \label{eq:Ap}
    \Ap(t)=
    \begin{cases}
      \qquad\qquad\qquad0 & \text{if } t<0 \\
      ~-\lambda\beta \bigl ( C_{AV}(0) - C_{AV}(t) \bigr )& \text{if } t>0 \\
    \end{cases}.
\end{equation}

\section{The Green-Kubo expression for transport coefficients}

We now consider an external potential, $\varv$, coupling to a conserved density, $n(\bm{r};\Gamma)$, as:
\begin{align}
    V(\Gamma) = \int \varv(\bm{r}) n(\bm{r};\Gamma) d\bm{r}. \label{eq:perturbation}
\end{align}
In the specific case of a dye dissolved into a solvent, illustrated in Fig. \ref{fig:Diffusion}, the relevant conserved density is the number density of dye molecules, whose phase-space expression can be assumed to be $n(\bm{r};\Gamma) = \sum_{n} \delta(\bm{r}-\bm{R}_{n})$, while the corresponding current is $\bm{\jmath} (\bm{r},\Gamma) = \sum_{n} \delta(\bm{r} - \bm{R}_{n}) \bm{R}_{n}/M_n$, $M_n$ being the mass of the $n$-th dye molecule, and the transport coefficient is proportional to the dye's diffusivity. Note that the phase-space expression of a conserved (current) density is not necessarily univocally defined. Different expressions that coincide in the long-wavelength limit sample the same hydrodynamical variable and give rise to the same transport coefficient. This is a consequence of the \emph{gauge invariance} of transport coefficients,\cite{Marcolongo2016,Ercole2016,Grasselli2021} which plays a fundamental role in transport theory.

When a classical observable is evaluated along a dynamical trajectory, $\Gamma_{t}$, the resulting function depends on time and on the initial conditions of the trajectory. Averaging the phase-space expression of a conserved density in the presence of an external perturbation with respect to the initial conditions of a perturbed molecular trajectory may result in a time-dependence of its expected value:
\begin{align}
    \begin{aligned}
        n(\bm{r},t) & =\langle n(\bm{r};\Gamma_{t}')\rangle_{0}\\
        & = \int n(\bm{r};\Gamma_{t}') \mathsf{P}_0 (\Gamma_{0})d\Gamma_{0}.
    \end{aligned} 
\end{align}

\subsection*{A hit-and-run (and slightly wrong) argument yielding the correct result}
If the perturbation, Eq. \eqref{eq:perturbation}, is switched on at time $t=0$, Eq. \eqref{eq:Ap} states that the flux measured at a later time, $t >0$, reads:
\begin{equation} \label{eq:J(T)a}
    \JJ(t) = \beta \int_\Omega \langle \hat \JJ(t) \hat n(\bm r) \rangle \varv(\bm r)d\bm r,
\end{equation}
where the equal-time correlation function in Eq. \eqref{eq:Ap} has been set to zero because current and density have opposite parities with respect to time reversal. By expressing $\hat n(\bm r)$ as the integral from zero to $t$ of its time derivative, using the continuity equation, Eq. \eqref{eq:continuity}, and integrating by parts with respect to $\bm r$, Eq. \eqref{eq:J(T)b} can be cast into the form:
\begin{equation} \label{eq:J(T)b}
    \JJ(t) = \beta \int_\Omega d\bm r\int_0^t dt' \langle \hat \JJ(t) \hat\jj(\bm r, t') \rangle \cdot \ff(\bm r), 
\end{equation}
where $\ff(\bm r)=-\nabla \varv(\bm r)$ is the external force acting on the system and the vector product within the brackets is an outer product (\emph{i.e.} the scalar product out of the expectation value is assumed to act between $\hat\jj(\bm r, t')$ and $\ff(\bm r)$). By assuming now that $\ff(\bm r)\approx \FF$ does not vary appreciably over the system's volume and that $t$ is larger than the current autocorrelation time, Eq. \eqref{eq:J(T)b} can be finally cast into the form of Eq. \eqref{eq:Onsager}, where
\begin{equation}
    \sigma = \frac{\Omega}{k_BT} \int_0^\infty \langle \hat {J}(t) \hat{J}(0) \rangle, \label{eq:GK}
\end{equation}
and $\hat J$ is any Cartesian component of $\hat\JJ$, which is the celebrated Green-Kubo expression for Onsager's transport coefficients.\cite{green1952,*green1954,kubo1957a,*kubo1957b}

According to Eqs. \eqref{eq:Onsager} and (\ref{eq:J(T)b}-\ref{eq:GK}), an external static force, which is even with respect to time inversion, induces a steady current, which is instead odd, thus apparently violating time-reversal symmetry. This should not come as surprise, because, according to Eq. \eqref{eq:SSource}, the response of the current to the external force determines the production of entropy, which is an intrinsically irreversible phenomenon. How come a microscopically reversible system (the Hamiltonian, Eq. \eqref{eq:unperturbed_H}, is indeed time-reversal invariant) gives rise to irreversible phenomena, such as diffusion, dissipation, and entropy production?

\subsection*{A better argument}

Physical systems are never infinite nor are external fields acting on them ever homogeneous. By the same token, no perturbation nor the response induced by it can ever be strictly time-independent. To make sense of the apparent breach of time-reversal symmetry, let us first consider the response of the system to a monochromatic perturbation. To this end, we first express the perturbation of Eq. \eqref{eq:perturbation} in terms of the Fourier components of the particle density and external potential. By leveraging Parseval's theorem,\cite{Parseval} the perturbation in Eq. \eqref{eq:perturbation} can be written as:
\begin{align}
    V(\Gamma,t) =  \Omega \sum_{\bm k} \tilde \varv(\bm k,t) {\tilde n}(-\bm k,\Gamma), \label{eq:perturbation-k}
\end{align}
where the sum extends to all the (quasi-discrete) wavevectors compatible with periodic boundary conditions, usually adopted in molecular simulations.

Let us suppose that a monochromatic static perturbation  of wavevector $\bm k$ is switched on at time $t=0$. Because of translational invariance, only the $\bm k$-th Fourier component of the (current) density will respond to the perturbation. According to Eq. \eqref{eq:Ap}, the current response to this perturbation is:
\begin{equation}  \label{eq:jCdot}
    \begin{aligned}
       \tilde\jj(\bm k,t)&=
       \beta\Omega\tilde \varv(\bm k) \langle \tilde\jj (\bm k,t) \tilde n(-\bm k,0) \rangle \\
                         &= -\beta\Omega \tilde \ff(\bm k) \frac{1}{k^2} \dot C(\bm k,t),
    \end{aligned}
\end{equation}
where $C(\bm k,t) = \langle {\tilde n}(\bm k,t) {\tilde n}(- \bm k,0)\rangle $ is the density-density correlation function, $\tilde \ff(\bm k)=-i{\bm k} \tilde \varv(\bm k)$ is the Fourier transform of the force acting on the system, and the continuity equation, Eq. \eqref{eq:continuity}, is used to express the (longitudinal component of the) current in terms of the density, $\tilde\jj(\bm k,t)=i\frac{\bm k}{k^2} \dot{\tilde n}(\bm k,t)$. In Eqs. \eqref{eq:jCdot} and thereafter a dot indicates a time derivative and all the carets indicating the phase-space dependence of the various Fourier coefficients have been suppressed to unburden the notation. According to Eq. \eqref{eq:jCdot}, the $t\to\infty$ limit of the current is zero, because density fluctuations become uncorrelated at large time lags, in agreement with our expectation that in the long-time limit the system must come to thermal equilibrium. Where do Eqs. \eqref{eq:Onsager} and (\ref{eq:J(T)b}-\ref{eq:GK}) then come from?

According to Onsager's regression hypothesis, Eq. \eqref{eq:regression}, the density autocorrelation function, $ C(\bm k,t)$, for positive times obeys the same time evolution as a small deviation of the density from equilibrium. In the hydrodynamic (long-time, long-wavelength) limit, where the diffusion equation for density fluctuations, Eq. \eqref{eq:diffusion}, holds, the density-density correlation function has the form of Eq. \eqref{eq:n(k,t)}:
\begin{equation} \label{eq:C(k,t)}
   C(\bm k,t) = C(\bm k,0) e^{-Dk^2t},
\end{equation}
where $C(\bm k,0) = \langle | \tilde n(\bm k) |^2 \rangle$. By inserting these relations into Eq. \eqref{eq:jCdot}, one obtains:
\begin{align}    
    \tilde \jj(\bm k,t)&= \sigma(\bm k)\tilde \ff(\bm k) e^{-Dk^2t}, \quad\text{where} \label{eq:j(k,t)a} \\
    \sigma(\bm k) &= \beta\Omega D \langle | \tilde n(\bm k) |^2 \rangle. \label{eq:sigma(k)}
\end{align}
Standard fluctuation theory\cite{Landau1996} indicates that the long-wavelength limit of the density fluctuations is proportional to the number/chemical-potential susceptibility in the grand canonical ensemble: $\lim_{\bm k\to 0} \langle | \tilde n(\bm k) |^2 \rangle = \langle \Delta N^2 \rangle/\Omega^2 = \frac{k_BT}{\Omega} \frac{\partial n}{\partial\mu}$. One sees that at any finite wavevector the large-time limit of the current vanishes, the decay time being longer, the larger the wavelength. However, at any given time the long-wavelength limit of the current is finite, and independent of time:
\begin{equation} \label{eq:j(0,r)a}
    \lim_{\bm k\to 0} \tilde \jj(\bm k,t) = \sigma\FF,
\end{equation}
where $ \sigma = \sigma(0) = D\frac{\partial n}{\partial \mu}$, which is the linear relation between currents and forces, Eq. \eqref{eq:Onsager}, we are after.

We observe that the apparent breach of time-reversal symmetry manifests in the dependence of the long-time (low frequency) and long-wavelength (low wavevector) limits on the order in which they are performed: when the former is performed first, the response displays the expected approach to equilibrium at any finite wavelength. In contrast, when the order of the limits is reversed, a steady current is apparently observed in response to a static, homogeneous ($\bm k\to 0$), perturbing force. While this argument is correct in the hydrodynamic limit, it may sound somewhat unsatisfactory, for the expression of the density correlation function on which it is based, Eq. \eqref{eq:C(k,t)}, is wrong at short times. In fact, $C(\bm k,t)$ is an even function of time. Therefore, when continued to negative times, $C(\bm k,t)=C(\bm k,0) e^{-Dk^2 |t|}$ would display an unphysical cusp at $t=0$, resulting in a finite current at short positive times, according to Eq. \eqref{eq:j(k,t)a}, and determining an unphysical discontinuity at $t=0$. Although this shortcoming would not undermine the validity of Eq. \eqref{eq:j(k,t)a} in the hydrodynamic regime, within which it has been obtained, it would be instructive to derive it from a model that smoothly bridges the regimes in which the current vanishes as $t\to 0^+$ and the hydrodynamic one, valid for $t\gtrsim 1/Dk^2$. The discontinuity of the current at short times can be traced back to the unphysical instantaneous response of the current to density gradients, assumed in the Fick's equation, Eq. \eqref{eq:fick}. We thus replace this equation with the more general linear constitutive relation:\cite{Forster}
\begin{subequations}\label{eq:memory-fick}
   \begin{align}
      \bm{\jmath}(\bm r,t)&=-\int_{0}^t D'(t-t') \nabla n(\bm r,t')dt',\label{eq:r-memory-fick} \\ 
      \tilde{\bm{\jmath}}(\bm k,t)&=-i \bm k\int_{0}^t D'(t-t') \tilde n(\bm k,t')dt', \label{eq:k-memory-fick}
   \end{align}
\end{subequations}
where $D'(t)$ is a diffusion memory kernel and the upper limit of integration is set to $t$ so as not to violate causality. In order to understand qualitatively the effects of diffusion memory on current dynamics, let us suppose that they are characterized by single relaxation time, $\tau$, so that the memory kernel can be written as:
\begin{align}
  D'(t)=\frac{D}{\tau}\mathrm{e}^{-\frac{t}{\tau}} \label{eq:D-memory-kernel}.
\end{align}
With this ansatz for the memory kernel, by combining equation \eqref{eq:k-memory-fick} with the continuity equation, Eq. \eqref{eq:continuity}, we obtain:
\begin{align}
  \tau \frac{\partial^2 \tilde n}{\partial t^2} + \frac{\partial \tilde n}{\partial t} +Dk^2 \tilde n = 0, \label{eq:GenDiff}
\end{align}
which, for $\tau=0$, is the Fourier transform of the diffusion equation, Eq. \eqref{eq:diffusion}, as it must.

The density-density correlation function, $C(\bm k,t)$ is the solution to Eq. \eqref{eq:GenDiff}, subject to the initial conditions $\tilde n(\bm k,0)=C(\bm k,0)$ and $\dot{\tilde n}(\bm k,0)=0$:
\begin{multline} \label{eq:C-memory}
    C(\bm k,t)= C(\bm k,0)e^{-\frac{t}{2 \tau }} \left[
    \sinh \left( \frac{t}{2\tau} \sqrt{1-4 D k^2 \tau }
        \right) \right .\\ \left .\Big / \sqrt{1-4 D k^2 \tau } \right .
        \left . + ~\cosh \left(\frac{t}{2\tau}
        \sqrt{1-4 D k^2 \tau } \right ) \right].
\end{multline}

\begin{figure}[t]
    \centering
    \includegraphics[width=0.87\columnwidth]{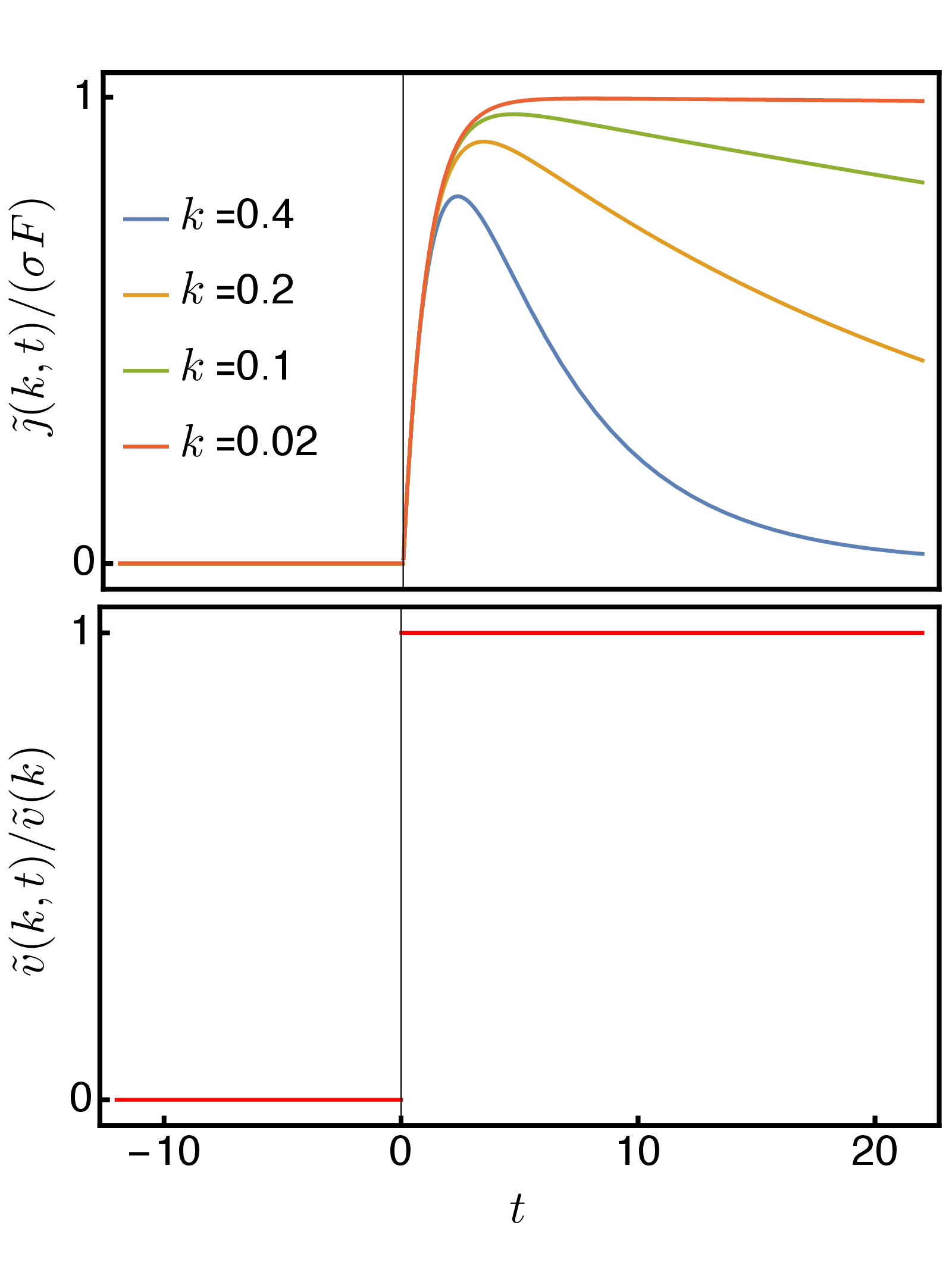} 
    \caption{Bottom: a monochromatic potential of wavenumber $\bm k$ and constant strength is switched on at $t=0$: $\tilde \varv(\bm k,t)= \tilde \varv(\bm k) \theta(t)$. Top: current response, $\bm\jmath(\bm k,t)$ to the perturbation depicted below, for different wavevectors of the perturbation. The units in Eq. \eqref{eq:GenDiff} are such that $D=1$ and $\tau=1$.} \label{fig:StepFunctPert}
\end{figure}

If a perturbation of wavevector $\bm k$ is switched on at time $t=0$, as depicted in in the bottom panel of Fig. \ref{fig:StepFunctPert}, and the system is then allowed to relax to equilibrium, the current response to the perturbation is given by Eq. \eqref{eq:jCdot}:
\begin{equation} \label{eq:j-memory}
   \tilde \jj (\bm k,t) = \sigma(\bm k) \tilde \ff(\bm k)  \frac{2 D e^{-\frac{t}{2\tau}}
   \sinh \left(\frac{t}{2\tau} \sqrt{1-4 D k^2 \tau} 
   \right)}{\sqrt{1-4 D k^2 \tau}}.
\end{equation}
Not surprisingly, Eqs. (\ref{eq:C-memory}-\ref{eq:j-memory}) reduce to Eqs. (\ref{eq:C(k,t)}-\ref{eq:sigma(k)}) in the hydrodynamic regime (large $t$, small $\bm k$). In particular, the $\bm k\to 0$ limit of the current in Eq. \eqref{eq:j-memory} is:
\begin{equation} \label{eq:j(0,t)b}
    \JJ(t) = \sigma \bm F (1-e^{-\frac{t}{\tau}}),
\end{equation}
which, in the large-time ($t\gg\tau$) limit reduces to Eq. \eqref{eq:j(0,r)a} and is therefore independent of time. The actual ($t\to \infty$, $\bm k\to 0$) limit depends on the order in which these limits are performed, as discussed before.

Fig. \ref{fig:StepFunctPert} displays the behavior of the current as a function of time for different values of the wavevector, $k=|\kk |$. Upon switching on the perturbation, the current rapidly rises from zero to a plateau value during a transient period determined by the relaxation time, $\tau$, which is independent of the wavevector. Once this plateau is reached, the system relaxes to a zero-current equilibrium state in a time that is longer the smaller the wavevector. This behavior is consistent with the hydrodynamic nature of the response of a conserved density to an external perturbation. In practice, observing the hydrodynamic decay of the current in a simulation requires increasing the size of the simulation cells to a value $\ell>\frac{2\pi}{k}$.

\subsection*{Transport coefficients from hydrodynamic susceptibilities}
The time dependence of the long-wavelength limit of the current, Eq. \eqref{eq:j(0,t)b}, has the general form:
\begin{equation}
    \JJ(t)=\sigma \FF + \bm I(t),
\end{equation}
where $\bm I(t)$ is integrable over the positive real axis: $\int_0^\infty \bm I(t) dt < +\infty$. The time Fourier transform of $J(t)$ is therefore:
\begin{equation}
    \tilde \JJ(\omega) = \int_0^\infty \FF(t) e^{i\omega t} dt =i\frac{\sigma \bm F}{\omega+i\epsilon} +\tilde {\bm I}(\omega), 
\end{equation}
where $\epsilon$ is an infinitesimal positive converging factor and $\tilde {\bm I}(\omega)$ is regular in the $\omega\to 0$ limit. We conclude therefore that
\begin{equation} \label{eq:sigmaFa}
    \sigma F = -i\lim_{\omega\to 0} \lim_{k\to 0} \omega \tilde {\tilde \jmath}(\bm k,\omega),
\end{equation}
where the double tilde indicates Fourier transform with respect to both space and time.

Let's now see what the result would be if the order of the limits in Eq. \eqref{eq:sigmaFa} were inverted. To this end, let us compute the Fourier transform of Eq. \eqref{eq:j-memory}:
\begin{equation}
    \begin{aligned}
        \tilde{\tilde \jj}(\bm k,\omega)
          &=\int_0^\infty \tilde\jj (\bm k,t) e^{i\omega t} dt \\
          &= i\frac{\sigma \tilde {\tilde {\bm f}}(\bm k,\omega)}{\omega +i Dk^2 -\omega^2\tau}
    \end{aligned}
\end{equation}
It is evident that, for any $\bm k\ne 0$,
\begin{equation}
    \lim_{\omega\to 0} \omega \tilde {\tilde \jj}(\bm k,\omega) = 0,
\end{equation}
and therefore the limits in Eqs. \eqref{eq:sigmaFa} do not commute.

In order to compute the transport coefficient in terms of the number susceptibility, $\tilde \chi(\kk,t-t')=\frac{\delta \tilde n(\kk,t)}{\delta\tilde\varv(\kk,t')}$, let us first express the current response to the perturbation depicted in Fig. \ref{fig:StepFunctPert} in term of it. By definition of the susceptibility, the density response reads:
\begin{equation}
    \tilde n(\bm k,t) = \int_{-\infty}^t \tilde \chi(\bm k,t-t') \tilde\varv(\kk,t')dt'.
\end{equation}
By expressing the potential in terms of the force, $\tilde \varv(\bm k, t) = -i \frac{{\bm k} \cdot \tilde {\bm f}(\bm k)}{k^2} \theta(t)$ and the (longitudinal) current in terms of the density through the continuity equation, $\tilde \jj(\bm k,t) = i \frac{\bm k}{k^2}\dot{\tilde n}(\kk,t)$, one obtains:
\begin{align}
    \tilde n(\bm k,t) &= -i \frac{{\bm k} \cdot \tilde {\bm f}(\bm k)}{k^2} \int_0^t \tilde\chi(\bm k, t-t') dt' \\ \tilde{\tilde \jj}(\bm k,\omega) &= \frac{1}{k^2} \tilde \ff(\bm k) \tilde{\tilde \chi}(\bm k,\omega).
\end{align}
By comparing these equations with Eq. \eqref{eq:sigmaFa}, we arrive at the final expression,
\begin{equation}
    \sigma = \lim_{\omega\to 0} \lim_{k\to 0} \frac{\omega}{k^2} \chi''(\bm k,\omega),
\end{equation}
where $\chi''= \mathsf{Im}~ \tilde{\tilde\chi}$ is the imaginary part of the density susceptibility, and the limits must be performed in the prescribed order, which is a frequently used alternate way of expressing the Green-Kubo relation, Eq. \eqref{eq:GK}.\cite{Kadanoff1963,Forster}

\section{Conclusions}
While not adding much to the well established Green-Kubo theory of irreversible processes,\cite{green1952,*green1954,kubo1957a,*kubo1957b} and its connection with the dynamics of hydrodynamic fluctuations and transport phenomena,\cite{Kadanoff1963,Forster} I hope that this note will help clarify the nature of the \emph{dynamical}, rather than \emph{static}, relationship between currents and thermodynamic forces. This relation is a characteristic feature of a transient state characterizing the approach to equilibrium, in the limit when the relaxation time is much larger than the observation time. While hardly a surprise, it is hoped that the examples and derivations presented in this note will shed some light onto a matter that is too often poorly explained in textbooks and whose proper understanding is given for granted in the classroom, as well as in the scientific literature. \bigskip

\begin{acknowledgments}
This paper is dedicated to Giovanni Ciccotti, one of the founding fathers of modern molecular dynamics, on the occasion of his eightieth birthday. Happy Birthday, Giovanni! This work is the offspring of an online course on transport theory that I gave with Federico Grasselli in the gloomy Spring of 2020, which, much to our honour and pride, was attended by Giovanni as well. I owe Federico good memories of the course and many insightful discussions on transport theory ever since. I am grateful to the anonymous referee of one of my recent works,\cite{Drigo2023} who objected to our cavalier inversion of the low-frequency / long-wavelength limits of a hydrodynamic susceptibility, thus providing the motivation to write the present paper. This work was partially supported by the European Commission through the \textsc{MaX} Centre of Excellence for supercomputing applications (grant number 101093374), by the Italian MUR, through the PRIN project  \emph{ARES} (grant number 2022W2BPCK) and by the Italian National Centre for HPC, Big Data, and Quantum Computing (grant number CN00000013), funded through the \emph{Next Generation EU} programme.
\end{acknowledgments}

\bibliography{GiovanniCiccotti}

\end{document}